# Thermalization dynamics in photonic lattices of different geometries

GUOWEN YANG[1,7], DOMENICO BONGIOVANNI[1,2,7], DAOHONG SONG[1,3], ROBERTO MORANDOTTI[2], ZHIGANG CHEN[1,3,6], AND NIKOLAOS K. EFREMIDIS[1,4,5,6]

[1]*MOE Key Laboratory of Weak-Light Nonlinear Photonics, TEDA Applied Physics Institute and School of Physics, Nankai University, Tianjin, 300457, China*
[2]*INRS-EMT, 1650 Blvd. Lionel-Boulet, Varennes, QC J3X 1S2, Canada*
[3]*Collaborative Innovation Center of Extreme Optics, Shanxi University, Taiyuan, Shanxi 030006, China*
[4]*Department of Mathematics and Applied Mathematics, University of Crete, Heraklion, Crete 70013, Greece*
[5]*Institute of Applied and Computational Mathematics, FORTH, Heraklion, Crete 70013, Greece*
[6]*Authors to whom correspondence should be addressed:*
*zgchen@nankai.edu.cn and nefrem@uoc.gr*
[7]*These authors contributed equally to this work*

**Abstract:** The statistical mechanical behavior of weakly nonlinear multimoded optical settings is attracting increased interest during the last few years. The main purpose of this work is to numerically investigate the main factors that affect the thermalization process in photonic lattices. In particular, we find that lattices with identically selected properties (such as temperature, coupling coefficient, lattice size, and excitation conditions) can exhibit very different thermalization dynamics and thus thermalization distances. Our investigation is focused on two different two-dimensional lattices: the honeycomb lattice and the triangular lattice. Our numerical results show that, independently of the excitation conditions, the honeycomb lattice always thermalizes faster than the triangular lattice. We mainly explain this behavior to the quasilinear spectrum that promotes wave-mixing in the honeycomb lattice in comparison to the power-like spectrum of the triangular lattice. In addition, we investigate the combined effects of temperature as well as the sign and magnitude of the nonlinearity. Switching either the sign of the Kerr nonlinear coefficient or the sign of the temperature can lead to significant differences in the thermalization dynamics, a phenomenon that can be physically explained in terms of wave instabilities. Larger absolute values of the temperature $|T|$ result in more uniform distributions for the power occupation numbers and faster thermalization speeds. Finally, as expected, increasing the magnitude of the nonlinearity results in accelerated thermalization. Our findings provide valuable insights into optical thermalization in discrete systems where experimental realization may bring about new possibilities for light manipulation and applications.

## 1. Introduction

Over the last few years, optical thermodynamics, the formulation of statistical mechanics for optical systems, has been established through extensive research efforts [1-3]. This theory was first put forward to explain the peculiar phenomenon of beam self-cleaning [4-6]. An intense light injected into a highly multi-mode optical parabolic graded-index fiber exhibits a bell-shaped output intensity pattern after nonlinear propagation. The beam self-cleaning phenomenon has been initially attributed to multi-wave mixing processes as an explanatory framework [7]. However, recent studies have introduced novel theoretical approaches based on thermodynamics, offering a more concise method for addressing and describing it, leading to the experimental observation of Rayleigh-Jeans (R-J) distribution in graded-index multimode fibers [8-10]. The emergence of the R-J distribution is directly linked to principles of statistical mechanics, in particular, the maximization of entropy within a multimode system with conserved quantities [1, 11-13]. Remarkably, it has been demonstrated that thermal equilibrium manifests across multiform multimode optical systems that obey certain conservation laws (i.e., the total optical power, and the internal energy), each of them characterized by distinct nonlinearities even beyond the multi-wave mixing paradigm [14]. In different optical systems, multiple theoretical studies have predicted various interesting thermal behaviors, such as entropic processes [15, 16], optical pressure in a multimoded optical setting [17], and also thermodynamics in topological and non-Hermitian structures [18, 19]. Further advances reporting the thermal equilibrium state with the additional conserved quantity of optical angular momentum (OAM) in fiber have been also investigated theoretically and experimentally [20, 21]. Also, the entropic behaviors under negative temperatures in both fiber and photonic lattices have been experimentally investigated [22-24].

In this work, we analyze the thermalization process and the associated thermalization distances in multimoded weakly nonlinear photonic lattices. In doing so, we utilize the Kullback-Leibler divergence (KLD), a measure that is additive for independent distributions [25-27]. In particular, we show that the thermalization processes between different lattice geometries can lead to thermalization distances that are significantly different. Specifically, here we select to compare honeycomb lattices with triangular lattices. To ensure a fair comparison, both lattices are chosen to have the same number of waveguides and the same nearest neighbor coupling coefficients. Furthermore, in both lattices, we excite the same number of consecutive modes with the same amount of power using a uniform distribution. These sequences of modes are selected to lead to approximately the same value of the optical temperature. We observe that even though both lattices have the same parameter set, there is a large difference in the thermalization dynamics with the honeycomb lattice being significantly faster. We attribute this difference to the spectrum of the honeycomb lattice that has a quasi-linear structure and thus promotes wave mixing phenomena. In addition, we examine the influence of different magnitudes and signs of optical temperature and nonlinearity on the thermalization speed of these two lattice geometries. Such a temperature depends on the spectral feature of lattice geometry and the excitation conditions, in particular power and internal energy, but is not related to the nonlinear strength. In general, larger absolute values of the temperature lead to more uniform distributions of the power occupation numbers and faster thermalization dynamics. Thus, it is significantly more difficult to observe thermalization close to the condensation limits $|T| \to 0$. In addition, wave instabilities affect the thermalization process leading to different thermalization distances, depending on the sign of $T \cdot \gamma$, where $\gamma$ is the Kerr nonlinear coefficient. Finally, as expected, the thermalization speed is increased by increasing the magnitude of the Kerr nonlinear coefficient. Our findings not only provide valuable insights into the intricate details of the thermalization process with a concise method but also offer practical guidance for the experimental realization of thermodynamic principles in lattice-based systems.

## 2. Thermodynamics of photonic lattices

In the presence of Kerr nonlinearity, paraxial light propagation in a photonic lattice structure is governed by the following nonlinear Schrödinger type equation

$$i\frac{\partial \psi}{\partial z} + \frac{1}{2k}\left(\frac{\partial^2 \psi}{\partial x^2} + \frac{\partial^2 \psi}{\partial y^2}\right) + V(x,y)\psi + \gamma|\psi|^2\psi = 0, \qquad (1)$$

where $\psi(x, y, z)$ is the electric field envelope, $x$ and $y$ are the transversal coordinates and $z$ the propagation distance, $V(x, y)$ is the lattice potential that is proportional to the refractive index, $\gamma$ is the Kerr nonlinear coefficient, and $k$ is the wavenumber. In particular, using coupled-mode theory, if next-nearest-neighbour couplings are negligible with respect to the coupling between the nearest neighbors, Eq. (1) can be simplified into the following discrete nonlinear Schrödinger equation

$$i\frac{d\psi_n}{dZ} + \sum_{\langle m,n \rangle} t\psi_m + \gamma|\psi_n|^2\psi_n = 0, \qquad (2)$$

where, in Eq. (2), $\psi_n$ is the complex field amplitude at the $n$th lattice site, Z represents the dimensionless propagation distance, and $t$ is the coupling coefficient between nearest neighbors $\langle m,n \rangle$. We expand the wavefunction $|\psi\rangle = \sum_i c_i|\psi_i\rangle$ into the eigenfunctions (or supermodes) $|\psi_i\rangle$ of the system, which satisfy

$$H|\psi_i\rangle = \varepsilon_i|\psi_i\rangle, \qquad (3)$$

where $H$ is the linear part of the Hamiltonian, $\varepsilon_i$ is the associated propagation constant of the $i$th mode, and, since each waveguide in isolation is a single mode, the total number of nodes or lattice sites $M$ is equal to the total number of modes. The power occupation number on each super mode is

$$n_i = |c_i|^2 = |\langle\psi_i|\psi\rangle|^2, \qquad (4)$$

and the total power $P = \langle\psi|\psi\rangle = \sum_i^M n_i$ is considered to be the first conservation law with the second being the internal energy approximately equal to its linear part $U = \langle\psi|H|\psi\rangle = \sum_i^M \varepsilon_i n_i$. Following the calculations, the optical entropy is found to satisfy [1]

$$S = \sum_{i=1}^M \ln n_i + M. \qquad (5)$$

In the case that the lattice is relatively large along both directions, the entropy $S = S(U, M, P)$ can be approximated to be extensive with respect to $(U, M, P)$ [1, 16]. The ensemble average value of the entropy increases in a spontaneous manner with the propagation distance $(d\langle S\rangle/dZ \geq 0)$ until asymptotically approaching its maximum value. In thermal equilibrium, the ensemble average power occupation on the eigenmodes obeys the Rayleigh-Jeans (R-J) distribution

$$\langle n_i \rangle = \frac{1}{\alpha + \beta\varepsilon_i} = \frac{T}{\varepsilon_i - \mu}, \qquad (6)$$

where $\alpha = -\mu/T$ and $\beta = 1/T$ denote the Lagrange multipliers of the constraints imposed by the conservation laws. This process of reaching thermal equilibrium is called thermalization, and $T$ and $\mu$ are defined as the optical temperature and chemical potential, respectively, being the conjugate variables to the internal energy and total power given by $T^{-1} = \partial S/\partial U$ and $\mu = -T\partial S/\partial P$. The extensive variables $U, P, M$ are related to $T, \mu$ through the following equation of the state [1]

$$U - \mu P = TM. \qquad (7)$$

For a given lattice system with eigenspectrum $\varepsilon_i$, which is excited with power $P$ and energy $U$, the optical temperature is uniquely determined by the expression

$$\sum_i^M \frac{T}{P\varepsilon_i - U + TM} = 1, \qquad (8)$$

while the chemical potential is then obtained from Eq. (7).

### 3. Thermalization dynamics

In the rest of this paper, we examine and compare the thermalization process of two photonic lattices under different conditions of the optical temperature and Kerr nonlinearity. Specifically, we focus our attention on investigating the thermalization processes associated with the honeycomb and the triangular lattices [see Figs. 1(a1,a2)]. Both of them belong to the same hexagonal Bravais group but have different features. In particular, the honeycomb lattice is a compound lattice structure with two atoms per unit cell, largely studied in optics due to a host of interesting topological phenomena it is associated with. Interestingly, it possesses a pair of energy bands that touch each other forming a Dirac-like cone, and such a singular band touching is demonstrated to induce the property of pseudospin around the Dirac points [28, 29]. On the other hand, the triangular lattice is one of the simplest two-dimensional lattice structures, with the unit cell containing a single atom. Figure 1(a1,a2) illustrates the schematic representations of the finite-sized honeycomb and triangular lattices examined here, where blue and red circles mark their respective lattice-site locations, while black lines connecting adjacent sites show the direction of the nearest neighbor couplings. The number of lattice sites for both lattices is set to $M = 144$, the nearest neighbor hopping amplitudes $t$ are taken to be 1, and zero boundary conditions are assumed throughout the analysis in order to consider a realistic scenario and account for boundary effects.

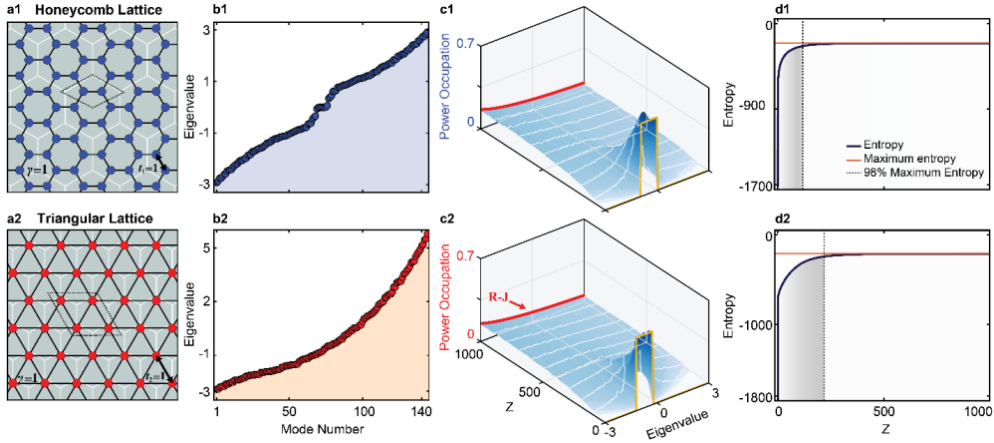

Fig. 1. (a1) Schematic illustration of a finite-sized honeycomb lattice with coupling coefficient $t = 1$, formed by $M = 144$ lattice sites. (b1) Corresponding real-space propagation constants. (c1) Ensemble average dynamics of the power occupation numbers illustrating the thermalization process occurring for Kerr-type nonlinearity with $\gamma = 1$ under uniform spectrum excitation of a band of supermodes (yellow line), and reaching thermal equilibrium as predicted by the Rayleigh-Jeans distribution (red line). (d1) Ensemble average value of the optical entropy (black curve) asymptotically approaching the predicted theoretical maximum value (red line), where the vertical dashed line marks the propagation distance where 98% of the maximum is reached. (a2-d2) Same as in (a1-d1) for a finite-sized triangular lattice with 144 nodes.

Under these conditions, the band structures are directly calculated by solving the eigenvalue problems of Eq. (3) and they are plotted in Fig. 1(b1,b2) as a function of the mode number. One can see that the energy spectrum associated with the finite-sized honeycomb lattice displays a quasi-straight-line profile as the mode number increases [Fig. 1b1], with the band-structure profile being symmetric with respect to zero energy level due to the chiral symmetry, which is preserved in this model. On the other hand, the band structure of the triangular counterpart [Fig. 1b2] exhibits a power-law-like relation with an exponent greater than that one for increasing mode-number values. Indeed, the band-structure profile of the triangular model is asymmetric with respect to the zero-energy level due to the lack of chiral symmetry [30]. In addition, numerical results in Fig. 1(c1,c2) compare the ensemble average thermalization process when light propagates in the two lattice structures in the presence of self-focusing nonlinearity. The initial conditions consist of a narrow band of eigenmodes that are uniformly excited (see yellow curves in Fig.1(c1,c2)). The numerical simulations are carried out by solving Eq. (2) using a Runge-Kutta numerical method. In particular, the total optical power is selected to be $P = 12.9$, and it is uniformly distributed among 21 consecutive eigenmodes, with a spectral bandwidth of $\Delta\varepsilon_{HC} = 0.01 - (-0.91) = 0.92$ for the honeycomb lattice and $\Delta\varepsilon_{TR} = -0.24 - (-1.16) = 0.92$ for the triangular lattice (see yellow lines in Fig. 1(c1,c2)) so that the resulting optical temperature at thermal equilibrium is the same for both lattices: $T = 0.5$. The power occupation numbers at each propagation distance $Z$ are averaged over 200 realizations. As shown in Fig. 1(c1,c2), during propagation, the ensemble average power occupations progressively evolve from the initial state towards the respective theoretically predicted R-J distributions (which are shown with the red curves in Fig. 1(c1,c2)). After a propagation distance $Z = 1000$ both systems are visually observed to reach their thermal equilibrium state. Looking at the ensemble average entropy curves, which are calculated from Eq. (5), as a function of the propagation distance $Z$, they increase and asymptotically approach the theoretically predicted thermal equilibrium value, as shown in Fig. 1(d1,d2). Furthermore, shaded gray areas enclosed between the entropy curves and the vertical dashed lines mark the locations where entropy reaches 98% of its maximum at the thermal equilibrium. Remarkably, they highlight the difference in the rates of changes of the entropies for both systems, thus indicating qualitatively that the thermalization process in the honeycomb lattice develops faster than the one occurring in the triangular system. In summary, our numerical results of Fig. 1 show that the honeycomb lattice tends to reach thermal equilibrium at shorter distances than the triangular lattice in comparable lattice and excitation conditions. A physical explanation can be found in the fact that the thermalization process is the outcome of different supermodes experiencing a nonlinear four-wave-mixing process, the latter of which is initiated by the presence of a small Kerr nonlinearity. The cause of different thermalization speeds in these two systems is attributable to the energy-spectral profile of their lattice geometries. In particular, honeycomb lattice features a quasi-straight-line profile of its energy spectrum or better a quasi-constant energy spacing, thereby optimizing the phase-matching conditions $k_1 + k_2 = k_3 + k_4$ and promoting the nonlinear wave-mixing phenomenon when compared with the triangular counterpart.

## 4. Kullback-Leibler Divergence (KLD)

In analyzing the thermalization dynamics, it is important to find a measure that directly compares the numerical ensemble average power occupation numbers with the theoretical predictions. In principle, we could use different normed metrics to measure the distance between these two distributions. However, we prefer to use the Kullback-Leibler divergence (KLD) (also known as the relative entropy), which is not formally a metric since it does not satisfy the symmetry relation $d(x, y) = d(y, x)$ or the triangle inequality. In particular, the relative entropy of the ensemble average power occupation numbers $\langle n_i \rangle$ at distance $Z$, from the theoretically predicted thermal equilibrium distribution $n_i^{RJ}$ is defined as

$$D_{KL}\left(n_i^{RJ}|\langle n_i(Z)\rangle\right) = D_{KL}(Z) = \sum_{i=1}^{M} \langle n_i(Z)\rangle \log\left(\frac{\langle n_i(Z)\rangle}{n_i^{RJ}}\right), \quad (9)$$

The KLD divergence is a useful measure in statistical mechanics that defines the distance between two distributions. It is a direct outcome of the Gibbs inequality, and as a result, it can only take non-negative values $D_{KL}(a|b) \geq 0$ and becomes zero $D_{KL}(a|b) = 0$ if and only if $a = b$. A property of the KLD, important in statistical mechanics, is that it is additive for independent distributions. An illustrative example where the $D_{KL}$ is plotted as a function of the propagation distance for the honeycomb lattice with $T = 0.5$ and $\gamma = 1$ is presented in Fig. 2. One can observe that the KLD has a decreasing trend as a function of the propagation distance, even though, especially for large propagation distances, it exhibits fluctuations. We have selected to set the thermalization threshold to the value $D_{KL} = 0.05$, which is supported by our numerical results. For example, in Fig. 2(b1)-(b4) we show the trend for the system to reach thermal equilibrium as the propagation distance increases. In particular, in Fig. 2(b1) we see the initial spectrum. At $Z = 300$, the system is still far from equilibrium [Fig. 2(b2)]. Although the higher eigenvalues are approaching the theoretically predicted distribution, the first eigenstates exhibit large deviations. In Fig. 2(b3) we show the power occupation numbers at the selected thermalization onset where $D_{KL} = 0.05$. We actually observe very good agreement between theory and numerical results. Finally, at $Z = 900$ (see Fig. 2(b4)) we have an even better agreement between the two distributions. For the purpose of confirming the high degree of similarity between the theoretical R-J distribution and the numerically calculated ensemble average power occupation numbers observed when $D_{KL} = 0.05$, we have statistically checked via a two-sample Kolmogorov-Smirnov test [31] that the confidence coefficient between the two curves is 0.99. Accordingly, we conclude that due to the high degree of similarity between the R-J and numerical distributions, the value $D_{KL} = 0.05$ can be considered as a thermal equilibrium threshold.

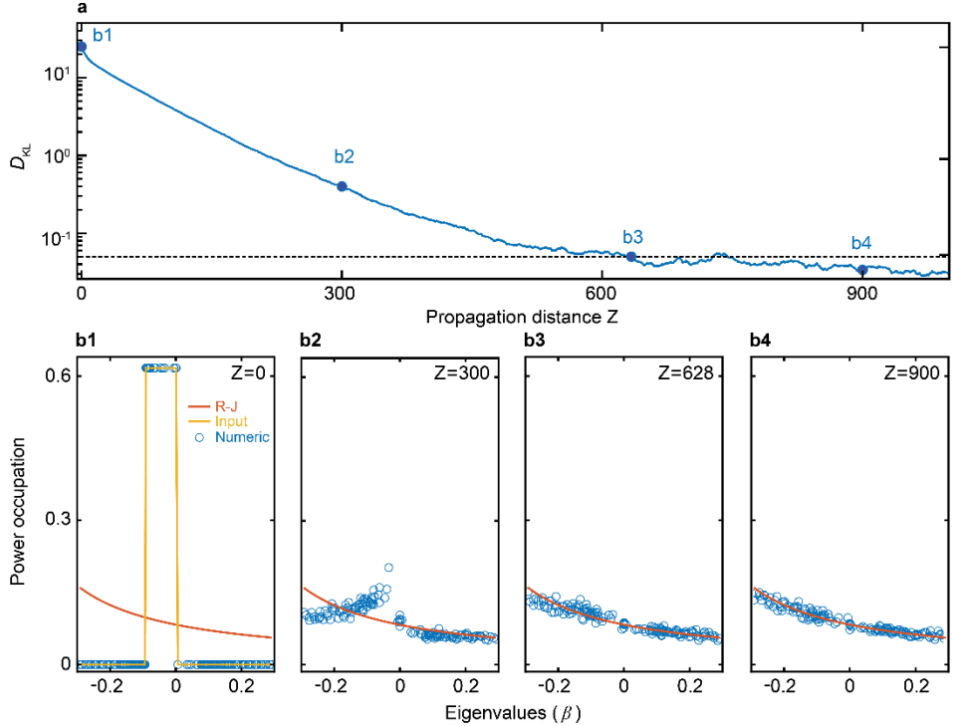

Fig. 2. (a) Relative entropy in a honeycomb lattice as a function of propagation distance Z. (b1-b4) Power occupations at selected Z under uniform initial excitation (yellow line) of the power $P = 12.9$ over 21 consecutive modes, compared with theoretical R-J distribution (red lines). The predicted value of the temperature is $T = 0.5$: (b1) $Z = 0$; (b2) $Z = 300$; (b3) $Z = 628$; and (b4) $Z = 900$. Power occupation curves approach the R-J distribution as $D_{KL}$ decreases below the 0.05 threshold value (dotted black line) at b3, displaying a confidence coefficient of 0.99 when estimated by a two-sample Kolmogorov-Smirnov test.

## 5. Thermalization dynamics in honeycomb and triangular lattices

Using as a main measure the mutual entropy, we will investigate the thermalization dynamics associated with photonic lattices under different temperature conditions and signs of the Kerr nonlinearity. As discussed in the previous section, it is reasonable to define the thermalization distance as the distance at which the mutual entropy $D_{KL}$ is equal to 0.05. First, we would like to examine whether different lattices under the same conditions have similar thermalization distances or not. In this respect, for simplicity, we are examining the two different types of lattice structures that are shown in Fig. 1(a): the honeycomb and the triangular lattices. To equalize their properties, both lattices have $M = 144$ nodes, coupling coefficient between first neighbors $t = 1$, and magnitude of the nonlinearity equal to $|\gamma| = 1$. In each configuration, the initial power is $P = 12.9$, which is uniformly distributed in a band of consecutive eigenmodes. This band is selected so that the predicted value of the temperature under thermal equilibrium is almost the same for both types of lattices. In Fig. 3(a), we present the thermalization distances for both triangular and honeycomb lattices, computed for five different positive and negative temperatures in the presence of self-focusing and self-defocusing nonlinearities ($|\gamma| = 1$). By comparing the thermalization speeds between the honeycomb and the triangular lattice, we see that under the same conditions (at least those presented in Fig. 3) the honeycomb lattice is always significantly faster than the triangular (note that the scale in the figure is logarithmic). We render this behavior to the quasi-linear type of spectrum of the honeycomb lattice that promotes power exchange between the modes. It is apparent from Fig. 3(a) that, independently of the sign and value of the optical temperature and the sign of nonlinearity, the thermalization process is accelerated as the absolute value of the optical temperature $|T|$ increases. Thus, it is significantly more difficult to achieve thermalization close to the condensation limits $|T| \to 0$. On the other hand, the configurations that are easier to achieve thermalization are associated with large absolute values of the temperature $|T|$, where the Rayleigh-Jeans distribution tends to become more uniform. The honeycomb lattice has a chiral symmetry and, thus, for each eigenmode $|\psi_i\rangle$ with eigenvalue $\varepsilon_i$, there exists an eigenmode $C|\psi_i\rangle$ with eigenvalue $-\varepsilon_i$, where $C$ is the chiral operator that changes the sign of one sublattice. The initial excitation conditions for opposite values of the temperature $T \to -T$ are related through the application of the chiral operator $|\psi\rangle \to C|\psi\rangle$. As a result, reversing both the temperature and the sign of the nonlinearity $\gamma \to -\gamma$ leaves the discrete nonlinear Schrödinger equation invariant. Thus, the ensemble average results obtained via the change of variable $(T, \gamma) \to (-T, -\gamma)$ should actually be identical. This is what we observe in Fig. 3(a), where the small fluctuations are the outcome of using a finite ensemble set. Note that such a symmetry is not present in the triangular lattice. As a result, the thermalization curves for positive and negative temperatures are different. In the case of the honeycomb lattice, it is interesting to observe that for the same absolute values of the temperature and nonlinearity, systems that satisfy the condition $T \cdot \gamma < 0$ will always thermalize faster. We can qualitatively understand this behavior in terms of modal instabilities. For example, in the case of positive temperature $T > 0$, the thermal equilibrium condition as described via the Rayleigh-Jeans distribution has larger power occupation numbers for the lower order modes. However, for $\gamma > 0$, the lower order modes (which are more "in-phase") are associated with instabilities and, thus, it is far more difficult to excite them in comparison to the $\gamma < 0$ regime where the lower order modes are more stable. The triangular lattice does not possess a chiral symmetry and its spectrum is power-like. We see that, on average, we achieve faster thermalization when the condition $T \cdot \gamma < 0$ is satisfied. However, approaching the condensation limit $T \to 0^-$ leads to far slower thermalization speeds as compared to the limit $T \to 0^+$. This can be physically explained by the form of the spectrum. Due to the power-law relation, it is significantly more difficult to excite the higher-order modes, which have larger spacing between them.

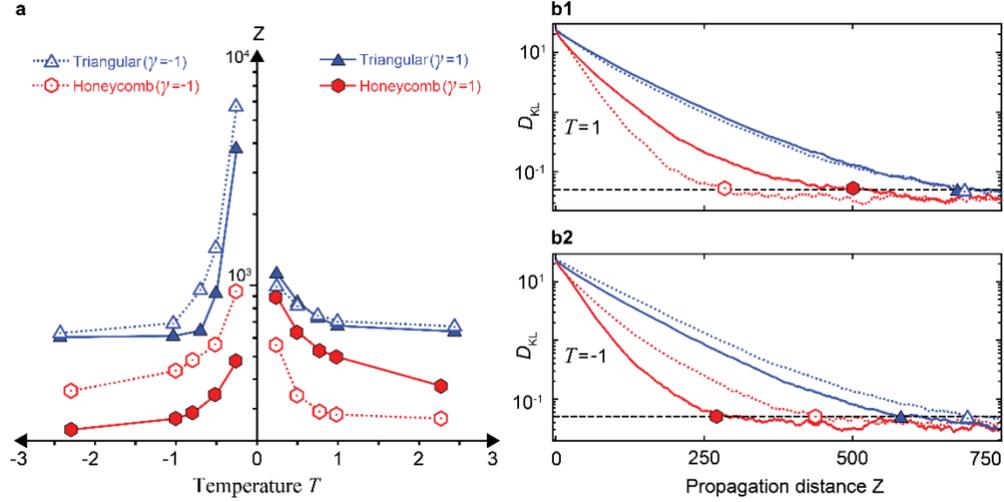

Fig. 3. (a) Thermalization distances $Z$ (in logarithmic scale) at different optical temperatures $T$ for self-focusing (solid) or self-defocusing (dotted) Kerr nonlinearity with $|\gamma| = 1$ in triangular (blue curves) and honeycomb (red curves) lattices with $M = 144$ modes and coupling coefficient $t = 1$. The initial excitation consists of 21 consecutive supermodes that are selected to match the same value of the optical temperature with total power $P = 12.9$. (b1, b2) Dynamics of the K-L divergence for (b1) $T = 1$ and (b2) $T = -1$, respectively. Dotted black lines in (b1) and (b2) mark the threshold value at 0.05.

Finally, in Fig. 3(b1,b2) we depict the dynamics of the relative entropy for light propagating in both honeycomb and triangular lattices under self-focusing and self-defocusing Kerr nonlinearities ($|\gamma| = 1$), for both positive and negative temperatures with $|T| = 1$. One can clearly see that under the same conditions, the $D_{KL}$ curves of the honeycomb lattice always cross the $D_{KL} = 0.05$ threshold lines earlier than the triangular counterpart, indicating faster thermalization. We see that in the honeycomb lattice with $T \cdot \gamma < 1$ thermalization is significantly faster than the rest of the combinations.

## 6. Effect of nonlinearity in the thermalization process

In this section, we investigate the effect of the strength of the nonlinearity in the thermalization process. We select to keep the same initial excitation and vary only the sign and value of the Kerr nonlinear coefficient $\gamma$. In our numerical simulations, the amount of nonlinearity is maintained small enough to avoid the formation of optical solitons during propagation, which will affect the thermalization process. Precisely, numerical simulations are performed for temperatures $T = 0.5$ by maintaining the same parameter set as in Fig. 3a, except for changing the strength of $\gamma$. As seen in Fig. 4(a), the thermalization process in the honeycomb lattice is observed to occur faster as compared to the triangular lattice for any sign and strength of the nonlinear coefficient $\gamma$. The main outcome is that increasing the strength of the nonlinearity $|\gamma|$ leads to a faster thermalization process. This is an expected result since, by using a modal decomposition, the wave mixing coefficients directly depend on $\gamma$.

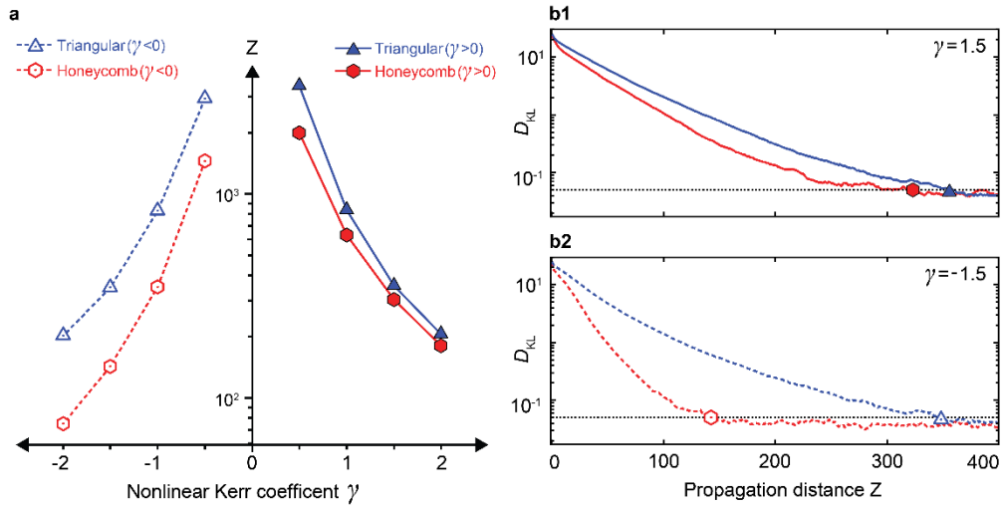

Fig. 4. (a) Thermalization distances Z (in logarithmic scale) for different Kerr nonlinear coefficients $\gamma$ in triangular (blue curves) and honeycomb (red curves) lattices, calculated at $T = 0.5$ using the same parameters as in Fig. 3(a). (b1, b2) Comparison of KLD dynamics for (b1) $\gamma = 1.5$ and (b2) $\gamma = -1.5$, respectively. Dotted black line in (b1) and (b2) marks the threshold value at 0.05.

In Fig. 4(b1, b2), typical $D_{KL}$ dynamics of the relative entropy are shown for $\gamma = 1.5$ and $-1.5$, respectively. The $D_{KL}$ curve of the honeycomb lattice for $\gamma = -1.5$ decreases faster and crosses the 0.05 line at an earlier distance as compared to the honeycomb lattice for the same temperature with $\gamma = 1.5$.

At the end of this section, we would like briefly to mention some ways to achieve thermalization dynamics in realistic physical realizations of triangular and honeycomb lattice structures. The main requirements for this purpose are a larger number of available supermodes and a waveguide length necessary to reach the thermal equilibrium. In photonic platforms, for example, lattice structures fabricated via femtosecond writing laser technique in glass or appropriate arrangements of coupled nonlinear single-mode fiber (SMF) could be employed. However, clear experimental observations of the thermalization phenomenon require an order of hundred coupling lengths.

## 7. Discussion and Conclusions

In conclusion, we have thoroughly investigated the thermalization dynamics occurring in honeycomb and triangular photonic lattices. Specifically, we have examined how the use of different lattice structures can affect the thermalization process, as well as the combined effects arising from the use of different temperatures, and signs and magnitudes of the Kerr nonlinearity. We would like to mention that the results of this work are focused in the case of honeycomb and triangular lattices. However, we expect that the results from this investigation may be applicable to other types of lattice structures, which certainly merit further investigation. However, we expect that our results will be particularly useful in this newly developed field of nonlinear optical thermodynamics, providing valuable insights into optical thermalization in discrete systems where experimental realization may bring about new possibilities for light manipulation and applications.

### Funding

We acknowledge financial support from the National Key R&D Program of China under Grant (2022YFA1404800), and the National Natural Science Foundation (12134006, 11922408, 12250410236). R. M. acknowledges support from NSERC Discovery and the CRC program in Canada. D.B. acknowledges support from the 66 Postdoctoral Science Grant of China.

### Disclosures

The authors declare no conflicts of interest.

### Data availability

Numerical results presented in this manuscript are available from authors upon reasonable request.